\def\ra{\rangle}
\def\la{\langle}
\def\be{\begin{equation}}
\def\ee{\end{equation}}
\def\ba{\begin{array}}
\def\ea{\end{array}}

\documentclass[aps,pre,preprint,showpacs,amssymb,amsfonts,preprintnumbers]
{revtex4}
\usepackage{epsfig,graphicx}
\usepackage{amsmath}
\usepackage{amscd,color}
\usepackage{soul}

\begin{document}

\baselineskip=18pt \setcounter{page}{1} \centerline{\large\bf
The norms of Bloch vectors and classification}\medskip
\centerline{\large\bf
 of four qudits quantum states} \vspace{4ex}
\begin{center}
Ming Li$^{1,2}$, Zong Wang$^{1}$, Jing Wang$^{1}$, Shuqian Shen$^{1}$ and Shao-ming Fei$^{2,3}$

\vspace{2ex}

\begin{minipage}{5in}

{\small $~^{1}$ College of the Science, China University of
Petroleum, 266580 Qingdao}

{\small $~^{2}$ Max-Planck-Institute for Mathematics in the
Sciences, 04103 Leipzig}

{\small $~^{3}$ School of Mathematical Sciences, Capital Normal University,
100048 Beijing}

\end{minipage}
\end{center}

\begin{center}
\begin{minipage}{5in}
\vspace{1ex} \centerline{\large Abstract} \vspace{1ex} We investigate the norms of the Bloch vectors for any quantum state with subsystems { less than or equal to} four. Tight upper bounds of the norms are { obtained}, which can be used to derive tight upper bounds for entanglement measure defined by the norms of Bloch vectors. { By using these bounds a} trade-off relation of the norms of Bloch vectors is discussed. Theses upper bounds are then applied on separability. Necessary conditions are presented for different kinds of separable states in four-partite quantum systems. { We further} present a complete classification of quantum states for four qudits quantum systems.

\vfill
\smallskip
\end{minipage}\end{center}
\bigskip

\section{Introduction}\label{sec1}
Quantum entanglement, as the remarkable nonlocal feature of quantum
mechanics, is recognized as a valuable resource in the rapidly
expanding field of quantum information science, with various
applications \cite{nielsen, di} such as quantum computation ,
quantum teleportation, dense coding, quantum cryptographic schemes,
quantum radar, entanglement swapping and remote states preparation. ${\rm {Tr}}$

It is known that the Bloch vectors give one
of the possible descriptions of qudit states. The Bloch vectors are then generalized to composite quantum systems with many subsystems.
From the norms of the Bloch vectors in the generalized Bloch
representation of a quantum state, separable conditions for both bi- and multi-partite quantum states
have been presented in \cite{vicente1,vicente2,hassan,ming}. Two multipartite
entanglement measures for N-qubit and N-qudit pure states are given in
\cite{hassan1,hassan2}. A general framework for detecting genuine multipartite entanglement
and non full separability in multipartite quantum systems of
arbitrary dimensions has been introduced in \cite{vicente3}.
In \cite{horo1995,mingbell} it has been shown that the norms of the Bloch vectors have
a close relationship to the maximal violation of a kind of multi Bell inequalities and to the concurrence \cite{GE,newge}. However, with the increasing of the dimensions of the subsystems, the norms of Bloch vectors for density matrices become hard to describe\cite{Mahler1995,Siennicki2001,Kimura2003,G.K2003}.

In this paper, we study the Bloch representations of quantum states with the number of subsystems less than { or equal to} four. We present tight upper bounds for the norms of Bloch vectors.
These upper bounds are then used to derive tight upper bounds for entanglement measure in \cite{hassan1,hassan2}. A trade-off relation of the norms of Bloch vectors is also discussed by { these} bounds.
Then we investigate different subclasses of bi-separable states in four-partite systems. Necessary conditions are presented for these kinds of separable states. By these analyses we present a complete classification of four qudits quantum states.

\section{Upper bounds of the norms of Bloch vectors}\label{sec2}

Let ${\lambda _i}$s $ ( {i = 1,...,{d^2} - 1}  )$ be orthogonal generators of $SU ( d  )$ which satisfy
$\lambda _i^+= {\lambda _i}, {\rm {Tr}}(\lambda _i) = 0,{\rm {Tr}}(\lambda _i \lambda _j) = 2{\delta _{ij}}.$
Denote the identity operator by ${{\rm I}_d}$. One finds that ${{\rm I}_d}$ and ${\lambda _i}$s compose an orthogonal basis of the linear space consisting of all $d\times d$ Hermitian matrices with respect to the Hilbert-Schmidt inner product. By using ${\rm {Tr}}\rho  = 1$ and $ \langle {{\lambda _i}}  \rangle  = {\rm {Tr}} ( {\rho {\lambda _i}}  )$ , we get that any density operator $\rho$ can be written in the form:
\begin{equation}\label{x}
\rho= \frac{1}{d}{{\rm I}_d} + \frac{1}{2}\sum\limits_{i = 1}^{{d^2} - 1} { \langle {{\lambda _i}}  \rangle } {\lambda _i}.
\end{equation}
The Bloch vector\cite{Bloch1946,Hioe1981,Pottinger1985,Lendi1986,Alicki1987,Mahler1995, G.K2003,Siennicki2001,Kimura2003}is defined by
${\bf{b}} =  ( {{b_1},...,{b_{{d^2} - 1}}}  ) \equiv  ( { \langle {{\lambda _1}}  \rangle ,..., \langle {{\lambda _{{d^2} - 1}}}  \rangle }  ).$
The state can be determined by { measuring} values of ${\lambda _i}$s, the state $\rho$ can also be given by the map
${\bf{b}} \to \rho  = \frac{1}{d}{{\rm I}_d}{\rm{ + }}\frac{{\rm{1}}}{{\rm{2}}}\sum\limits_{i = 1}^{{d^2} - 1} {{b_i}{\lambda _i}}.$
The set of all the Bloch vectors
that constitute a density operator is known as the Bloch vector
space $B({\Bbb{R}}^{d^{2}-1})$.

A matrix of the form (\ref{x}) is of unit trace and Hermitian, but
it might not be positive. To guarantee the positivity restrictions
must be imposed on the Bloch vector. It is shown that
$B({\Bbb{R}}^{d^{2}-1})$ is a subset of the ball
$D_{R}({\Bbb{R}}^{d^{2}-1})$ of radius
$R=\sqrt{2(1-\frac{1}{d})}$, which is the minimum ball
containing it, and that the ball $D_{r}({\Bbb{R}}^{d^{2}-1})$ of
radius $r=\sqrt{\frac{2}{d(d-1)}}$ is included in
$B({\Bbb{R}}^{d^{2}-1})$ {\cite{Harriman}}, that is,
$D_{r}({\Bbb{R}}^{d^{2}-1})\subseteq B({\Bbb{R}}^{d^{2}-1})\subseteq D_{R}({\Bbb{R}}^{d^{2}-1}).$

Using the generators of $SU(d)$, any quantum state $\rho\in
H_1^{d}\otimes H_2^{d}$ can be writing as: \be\label{1}
\rho=\frac{1}{d^2}I\otimes
I+\frac{1}{2d}\sum_{k=1}^{d^2-1}r_k\lambda_k\otimes I+\frac{1}{2d}\sum_{l=1}^{d^2-1}s_l
I\otimes\lambda_l+\frac{1}{4}\sum_{k=1}^{d^2-1}\sum_{l=1}^{d^2-1}t_{kl}\lambda_k\otimes
\lambda_l, \ee where $r_k={\rm {Tr}}(\rho\lambda_k\otimes I),
s_l={\rm {Tr}}(\rho I\otimes\lambda_l)$ and
$t_{kl}={\rm {Tr}}(\rho\lambda_k\otimes\lambda_l)$. We denote by
$T^{(12)}$ a vector with entries $t_{kl}$.
By using ${\rm {Tr}}(\rho^2)\leq 1$, one obtains that
\begin{equation}
 \|T^{(12)} \|^2 \le \frac{4(d^2-1)}{d^2},
\end{equation}
where $||\cdot||$ stands for the Hilbert-Schmidt norm or Frobenius norm.

We then consider the upper bounds of the Hilbert-Schmidt norm of the Bloch vectors for tripartite quantum systems.
Let $\rho  \in {\rm H}_1^d \otimes {\rm H}_2^d \otimes {\rm H}_3^d$ be a quantum state, which can be represented by Bloch vectors as follow:
\begin{equation}\label{s3}
\begin{aligned}
\rho  &= \frac{1}{{{d^3}}}I \otimes I \otimes I + \frac{1}{{2d^2}}(\sum {t_i^1{\lambda _i} \otimes I \otimes I}  + \sum {t_j^2I \otimes {\lambda _j} \otimes I}  + \sum {t_k^3I \otimes I \otimes {\lambda _k}} ) \\
&+ \frac{1}{{4d}}(\sum {t_{ij}^{12}{\lambda _i} \otimes {\lambda _j} \otimes I}        + \sum {t_{ik}^{13}{\lambda _i} \otimes I \otimes {\lambda _k}}  + \sum {t_{jk}^{23}I \otimes {\lambda _j} \otimes {\lambda _k}} )\\
&+ \frac{1}{8}\sum {t_{ijk}^{123}{\lambda _i} \otimes {\lambda _j} \otimes {\lambda _k},}   \\
\end{aligned}
\end{equation}
{ where $t_i^1={\rm {Tr}} ( {\rho {\lambda _i}
\otimes I \otimes I}  ), t_j^2={\rm {Tr}} ( {\rho I
\otimes {\lambda _j} \otimes I}  ), t_k^3={\rm {Tr}} (
{\rho I \otimes I \otimes {\lambda _k}}  ),
t_{ij}^{{\rm{12}}}={\rm {Tr}} ( {\rho {\lambda _i} \otimes {\lambda _j} \otimes I}  ),$ $t_{ik}^{{\rm{13}}}={\rm {Tr}} ( {\rho {\lambda _i} \otimes I \otimes {\lambda _k} }  ), t_{jk}^{{\rm{23}}}={\rm {Tr}} ( {\rho I \otimes {\lambda _j} \otimes {\lambda _k} }  ),
t_{ijk}^{{\rm{123}}}={\rm {Tr}} ( {\rho {\lambda _i} \otimes {\lambda _j} \otimes {\lambda _k}} )$ in the above representation. Define further ${T^{ ( x  )}}, {T^{ ( {xy}  )}}, {T^{ ( {123} )}}$ be the vectors with entries $t_i^x,t_{ij}^{xy},t_{ijk}^{123}$, and $1 \le x < y < z \le 3$.}

\textbf{Theorem 1:} For $\rho  \in {\rm H}_1^d \otimes {\rm H}_2^d \otimes {\rm H}_3^d$ with Bloch representation $(\ref{s3}),$ we have:
\begin{equation}
{ \| {{T^{ ( {123}  )}}}  \|^2} \le \frac{1}{{{d^3}}} ( {8{d^3} - 24d + 16}  )
\end{equation}

See supplemental material for the proof of the theorem.

We further consider four-partite quantum states. Let $\rho  \in {{\rm H}_{1234}}{\rm{ = H}}_1^d \otimes {\rm{H}}_2^d \otimes {\rm{H}}_3^d \otimes {\rm{H}}_4^d$ be a mixed quantum state with the Bloch representation
\begin{equation}\label{rho4}
\rho  = \frac{1}{{{d^4}}}I \otimes I \otimes I \otimes I+\frac{1}{{2{d^3}}}{M_1}+ \frac{1}{{4{d^2}}}{M_2}+\frac{1}{{8d}}{M_3}+\frac{1}{{16}}{M_4},
\end{equation}
where
\begin{eqnarray*}
{M_1}&=&\sum\limits_i {t_i^1{\lambda _i} \otimes I \otimes I \otimes I}+\sum\limits_j {t_j^2I \otimes {\lambda _j} \otimes I \otimes I}+\sum\limits_k {t_k^3I \otimes I \otimes {\lambda _k} \otimes I}\\
&&+\sum\limits_l {t_l^4I \otimes I \otimes I \otimes {\lambda _l}},\\
{M_2}&=&\sum\limits_{i,j} {t_{ij}^{12}{\lambda _i} \otimes {\lambda _j} \otimes I \otimes I}+ \sum\limits_{i,k} {t_{ik}^{13}{\lambda _i} \otimes I \otimes {\lambda _k} \otimes I}
+\sum\limits_{i,l} {t_{il}^{14}{\lambda _i} \otimes I \otimes I \otimes {\lambda _l}}\\
&&+\sum\limits_{j,k} {t_{jk}^{23}I \otimes {\lambda _j} \otimes {\lambda _k} \otimes I}+ \sum\limits_{j,l} {t_{jl}^{24}I \otimes {\lambda _j} \otimes I \otimes {\lambda _l}}+ \sum\limits_{k,l} {t_{kl}^{34}I \otimes I \otimes {\lambda _k} \otimes {\lambda _l}},\\
{M_3}&=&\sum\limits_{i,j,k} {t_{ijk}^{123}{\lambda _i} \otimes {\lambda _j} \otimes {\lambda _k} \otimes I}+\sum\limits_{i,j,l} {t_{ijl}^{124}{\lambda _i} \otimes {\lambda _j} \otimes I \otimes {\lambda _l}}+\sum\limits_{j,k,l} {t_{jkl}^{234}I \otimes {\lambda _j} \otimes {\lambda _k} \otimes {\lambda _l}},\\
{M_4}&=&\sum\limits_{i,j,k,l} {t_{ijkl}^{1234}{\lambda _i} \otimes {\lambda _j} \otimes {\lambda _k} \otimes {\lambda _l}}.
\end{eqnarray*}
We have defined $t_i^1={\rm {Tr}} ( {\rho {\lambda _i}
\otimes I \otimes I \otimes I}  ),t_j^2={\rm {Tr}} ( {\rho I
\otimes {\lambda _j} \otimes I \otimes I}  ),t_k^3={\rm {Tr}} (
{\rho I \otimes I \otimes {\lambda _k} \otimes I}  )$,
$t_l^4={\rm {Tr}} ( {\rho I \otimes I \otimes I \otimes {\lambda _l}}  )$,
$t_{ij}^{{\rm{12}}}={\rm {Tr}} ( {\rho {\lambda _i} \otimes {\lambda _j} \otimes I \otimes I} ),t_{ik}^{{\rm{13}}}={\rm {Tr}} ( {\rho {\lambda _i} \otimes I \otimes {\lambda _k} \otimes I} ),t_{il}^{{\rm{14}}}={\rm {Tr}} ( {\rho {\lambda _i} \otimes I \otimes I \otimes {\lambda _l}} )$,$t_{jk}^{{\rm{23}}}={\rm {Tr}} ( {\rho I \otimes {\lambda _j} \otimes {\lambda _k} \otimes I} ),t_{jl}^{{\rm{24}}}={\rm {Tr}} ( {\rho I \otimes {\lambda _j} \otimes I \otimes {\lambda _l}} ),t_{kl}^{34}={\rm {Tr}} ( {\rho I \otimes I \otimes {\lambda _k} \otimes {\lambda _l}}  )$,
$t_{ijk}^{{\rm{123}}}={\rm {Tr}} ( {\rho {\lambda _i} \otimes {\lambda _j} \otimes {\lambda _k} \otimes I}  ),t_{ijl}^{{\rm{124}}}={\rm {Tr}} ( {\rho {\lambda _i} \otimes {\lambda _j} \otimes I \otimes {\lambda _l}}  ),t_{ikl}^{{\rm{134}}}={\rm {Tr}} ( {\rho {\lambda _i} \otimes I \otimes {\lambda _k} \otimes {\lambda _l}}  )$,$t_{jkl}^{{\rm{234}}}={\rm {Tr}} ( {\rho I \otimes {\lambda _j} \otimes {\lambda _k} \otimes {\lambda _l}}  )$,
$t_{ijkl}^{{\rm{1234}}}={\rm {Tr}} ( {\rho {\lambda _i} \otimes {\lambda _j} \otimes {\lambda _k} \otimes {\lambda _l}}  )$ in the above representation. Define further ${T^{ ( x  )}},{T^{ ( {xy} )}},{T^{ ( {xyz}  )}},{T^{ ( {1234}  )}}$ be the vectors with entries $t_i^x,t_{ij}^{xy},t_{ijk}^{xyz},t_{ijkl}^{1234}$, and $1 \le x < y < z \le 4$.

\textbf{Theorem 2:} For $\rho  \in {\rm H}_1^d \otimes {\rm H}_2^d \otimes {\rm H}_3^d\otimes {\rm H}_4^d$ with Bloch representation $(\ref{rho4}),$ we have:
\begin{equation}
\|T^{(1234)}\|^2\le \frac{16(d^2-1)^2}{d^4}.
\end{equation}

See supplemental material for the proof of the theorem.

The two upper bounds for norms of Bloch vectors are tight and useful as that will be shown
in the following remarks.

\textbf{Remark 1:}
The Bloch vectors are used to define a valid entanglement measure in \cite{hassan1,hassan2} as follows. For a N-qudit pure
state, the entanglement measure is defined as
\be
E_{T}(|\psi\ra)=\frac{d^N}{2^N}||T^{(N)}||-(\frac{d(d-1)}{2})^{\frac{N}{2}},\ee
where $T^{(N)}$ is defined as a tensor with elements $t_{i_1i_2\cdots i_N}^{{\rm{12\cdots N}}}={\rm {Tr}} ( {\rho \lambda_{i_1} \otimes \lambda_{i_2}  \otimes \cdots \otimes \lambda_{i_N} }  )$.

By Theorem 1 and 2, one obtains the upper bounds of $E_{T}(|\psi\ra)$ for $N=3$ and $N=4$ as follows.
\be
E_{T}(|\psi\ra)\leq\left\{
\begin{aligned}
\sqrt{\frac{d^3(d-1)^2}{8}}(\sqrt{d+2}-\sqrt{d-1}),\ \ \ \ \ \ \ \ \ \ \ \ \ \ \ \ \ \textrm{N=3;}\\
\frac{d^2(d-1)}{2},\ \ \ \ \ \ \ \ \ \ \ \ \ \ \ \ \ \textrm{N=4.}
\end{aligned}\right.
\ee

By considering the the tripartite-qutrit state
$|\psi\ra=\frac{1}{\sqrt{3}}(|000\ra+|111\ra+|222\ra)$ and the four-qubit state
$|\phi\ra=\frac{1}{\sqrt{2}}(|0000\ra+|1111\ra)$, one computes the upper bounds of $E_{T}(|\psi\ra)$ are 3.01969 and 2 respectively(coincide with that in \cite{hassan2}). Thus the upper bounds of $E_{T}(|\psi\ra)$ are tight.

\textbf{Remark 2:} We consider four-partite quantum systems.
In \cite{zong} we have shown that for the state $\rho $ with representation (\ref{rho4}), we have:
\begin{equation}
\sum\limits_{1 \le x < y < z \le 4} {{{ \| {{{\rm{T}}^{( {xyz} )}}}  \|}^2}}  \le \frac{{8{{( {{d^2} - 1} )}^3}}}{{{d^3}({d^2} - 2)}},\label{az}
\end{equation}
where $||\cdot||$ stands for the $l_2$ norm of a vector.

Set $d=2$, we get $\sum\limits_{1 \le x < y < z \le 4}  \| \rm{T}^{( xyz )}  \|^2  \le 13.5$. By theorem 1 one has $ \| \rm{T}^{( xyz )}  \|^2\le 4$. Thus we obtain that it is impossible for  $\| \rm{T}^{(123)}\|, \| \rm{T}^{(124)}\|, \| \rm{T}^{(134)}\|,$ and $\| \rm{T}^{(234)}\|$ { attaining} 4 simultaneously.

\section{Necessary conditions for bi-separable states.}

In this section, we investigate subclasses of the bi-separable states in four partite quantum systems by the upper bounds of norms for Bloch vectors. Let's start with the following definition.

\textbf{Definition:}
Let $\rho\in {\rm H}_{\rm{1}}^d \otimes {\rm H}_2^d \otimes {\rm H}_3^d\otimes {\rm H}_4^d$ { be a quantum state with $d$ being} the dimension of the subsystems ${\rm H}_i, i=1,2,3,4.$
If $\rho$ can be written as
$\rho=\sum_kp_k|x_k\ra\la x_k|,$
where $\sum_kp_k=1, |x_k\ra$ is in one of the following sets:
$\{|\phi_1\ra \otimes |\phi_{234}\ra, |\phi_2\ra \otimes |\phi_{134}\ra, |\phi_3\ra \otimes |\phi_{124}\ra, |\phi_4\ra \otimes |\phi_{123}\ra\}$,
$\{|\psi_{12}\ra \otimes |\psi_{34}\ra, |\psi_{13}\ra \otimes |\psi_{24}\ra, |\psi_{14}\ra \otimes |\psi_{23}\ra\}$,
$\{|\xi_1\ra\otimes |\xi_2\ra \otimes |\xi_{34}\ra, |\xi_1\ra\otimes |\xi_3\ra \otimes |\xi_{24}\ra, |\xi_1\ra\otimes |\xi_4\ra \otimes |\xi_{23}\ra, |\xi_{14}\ra\otimes |\xi_2\ra \otimes |\xi_3\ra, |\xi_{13}\ra\otimes |\xi_2\ra \otimes |\xi_4\ra, |\xi_{12}\ra\otimes |\xi_3\ra \otimes |\xi_4\ra\}$
and $\{|\chi_1\ra \otimes |\chi_2\ra\otimes |\chi_3\ra\otimes |\chi_4\ra\}$, then $\rho$ is called $1-3$ separable, $2-2$ separable, $1-1-2$ separable, and $1-1-1-1$ separable respectively.

The following theorem gives necessary conditions of these kinds of separable states.

\textbf{Theorem 3:} Let $\rho  \in {\rm H}_1^d \otimes {\rm H}_2^d \otimes {\rm H}_3^d\otimes {\rm H}_4^d$ be a four-qudit quantum state. We have
$$ \| T^{(1234)}  \|^2\le\left\{
\begin{aligned}
\frac{16}{d^4}((d-1)(d^3-3d+2)),&&\textrm{if $\rho$ is 1-3 separable;}\\
\frac{16}{d^4}(d^2-1)^2,\ \ \ \ \ \ \ \ \ \ \ \ \ \ \ \ \ &&\textrm{if $\rho$ is 2-2 separable;}\\
\frac{16}{d^4}(d^2-1)(d-1)^2,\ \ \ \ \ \ \ \ &&\textrm{if $\rho$ is 1-1-2 separable;}\\
\frac{16}{d^4}(d-1)^4,\ \ \ \ \ \ \ \ \ \ \ \ \ \ \ \ \ \ \ &&\textrm{if $\rho$ is 1-1-1-1 separable}.
\end{aligned}
\right.$$

See supplemental material for the proof of the theorem.

The following two examples show that the upper bounds in theorem 3 are nontrivial and are tight.

\textbf{Example 1:} Consider the quantum state $\rho\in {\rm H}_1^d \otimes {\rm H}_2^d \otimes {\rm H}_3^d\otimes {\rm H}_4^d$,
\be
\rho=x|\psi\ra\la\psi|+\frac{1-x}{16}I,
\ee
where $|\psi\ra=\frac{1}{\sqrt{2}}(|0000\ra+|1111\ra)$ and $I$ stands for the identity operator.
By theorem 3, we compute that $\| T^{(1234)}\|^2=9x^2$. { Thus for $\frac{2}{3}<x\leq 1$ and $\frac{1}{\sqrt{3}}<x\leq \frac{2}{3}$, $\rho$ will be not 1-3 separable and not 1-1-2 separable respectively.} While for $\frac{1}{3}<x\leq \frac{1}{\sqrt{3}}$, $\rho$ is not 1-1-1-1 separable.

\textbf{Example 2:} Consider bi-separable state $\rho^d=|\psi_+^d\ra\la\psi_+^d|$ with $|\psi_+^d\ra=\frac{1}{\sqrt{d}}\sum_{i=1}^d|ii\ra\otimes\frac{1}{\sqrt{d}}\sum_{i=1}^d|ii\ra$.
One computes that $\| T^{(1234)}\|^2=\frac{16}{d^4}(d^2-1)^2$ which means that the upper bound for 2-2 separable states in theorem 3 is saturated. Actually, the upper bound can be also attained by considering the maximal entangled states as shown in remark 1.

\textbf{Remark 3:} With above theorems and examples, we are ready to classify the four-partite quantum states by using the norms of the Bloch vector $\| T^{(1234)}\|$, as shown in Fig.1.
It is worth mentioning that the 1-3 separable quantum states are always in the interior of the bi-separable set, while for some 2-2 separable quantum states the boundary of the bi-separable set is attainable. Since the upper bound for 2-2 separable states is just the upper bound for any four qudits states, we conclude that it is possible that the 2-2 separable state is on the boundary of the set of states(see Fig. 1).
\begin{figure}[h]
\begin{center}
\resizebox{7cm}{!}{\includegraphics{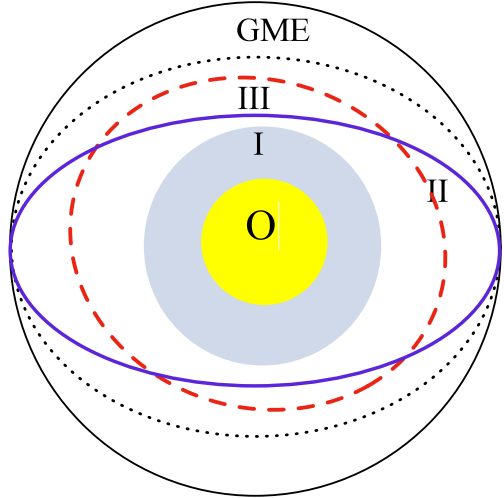}}
\end{center}
\caption{We put all the four-partite quantum states into a set(the largest circle). Quantum states located in the area between the largest
circle and the dotted oval are genuine multipartite entangled(GME). The bipartite
separable states are classified to 1-1-1-1 separable part(O, minimal and yellow circle), 1-1-2 separable part(I, blue area), 2-2 separable part(II, blue and solid oval), 1-3 separable part(III, red and dashed circle) and the rest part of the dotted oval.
\label{fig1}}
\end{figure}

\section{Conclusions and Remarks}\label{sec5}

It is a basic and fundamental question in quantum entanglement
theory to classify and detect entanglement states. In this paper, we have investigated the { norms} of the Bloch vectors for any quantum state with subsystems less than { or equal to} four. Tight upper bounds of the norms have been derived, which are used to derive tight upper bounds for entanglement measure defined by the norms of Bloch vectors. A trade-off relation of the norms of Bloch vectors is also discussed by { these} bounds. Then these upper bounds have been applied on the separability. Necessary conditions have been presented for 1-3, 2-2, 1-1-2 and 1-1-1-1 separable quantum states in four-partite quantum systems. With these bounds a complete classification of four qudits quantum states is presented.

\bigskip
\noindent{\bf Acknowledgments}\, \, This work is supported by the NSFC No.11775306, and 11701568; the Fundamental Research Funds for the Central Universities Grants No.16CX02049A, 17CX02033A and 18CX02023A; the Shandong Provincial Natural Science Foundation No.ZR2016AQ06, and ZR2017BA019.

\newpage
\title{Supplemental material for ``The norms of Bloch vectors and classification of four qudits quantum states"}
\author{Ming Li}
\affiliation{College of the Science, China University of Petroleum,
266580 Qingdao, China}
\affiliation{ Max-Planck-Institute for
Mathematics in the Sciences, 04103 Leipzig, Germany}
\author{Zong Wang}
\affiliation{College of the Science, China University of Petroleum,
266580 Qingdao, China}
\author{Jing Wang}
\affiliation{College of the Science, China University of Petroleum,
266580 Qingdao, China}
\author{Shuqian Shen}
\affiliation{College of the Science, China University of Petroleum,
266580 Qingdao, China}
\author{Shao-ming Fei}
\affiliation{ Max-Planck-Institute for
Mathematics in the Sciences, 04103 Leipzig, Germany}
\affiliation{School of Mathematical Sciences, Capital Normal University,
100048 Beijing, China}

\maketitle

\setcounter{equation}{0}
\renewcommand{\theequation}{S\arabic{equation}}

\subsection{Proof of Theorem 1}

{\bf{Proof:}} We start with the pure state. For an arbitrary pure state $\rho  = \left| \psi  \right\rangle \left\langle \psi  \right|$ one has ${\rm {Tr}}{\rho ^2} = 1,$ which means
\begin{eqnarray}\label{da}
{\rm {Tr}}{\rho ^2} &=& \frac{1}{{{d^3}}} + \frac{1}{{2{d^2}}}\left[ {\sum {{{(t_i^1)}^2} + } \sum {{{(t_j^2)}^2} + \sum {{{(t_k^3)}^2}} } } \right]\nonumber\\
&& + \frac{1}{{4d}}\left[ {\sum {{{(t_{ij}^{12})}^2} + \sum {{{(t_{ik}^{13})}^2} + \sum {{{(t_{jk}^{23})}^2}} } } } \right] + \frac{1}{8}\sum {{{(t_{ijk}^{123})}^2}}  = 1.
\end{eqnarray}
Set ${\left\| {{T^{(1)}}} \right\|^2} = {\sum {(t_i^1)} ^2}, {\left\| {{T^{(2)}}} \right\|^2} = {\sum {(t_j^2)} ^2}, {\left\| {{T^{(3)}}} \right\|^2} = {\sum {(t_k^3)} ^2}, {\left\| {{T^{(12)}}} \right\|^2} = {\sum {(t_{ij}^{12})} ^2}, {\left\| {{T^{(13)}}} \right\|^2} = {\sum {(t_{ij}^{13})} ^2}, {\left\| {{T^{(23)}}} \right\|^2} = {\sum {(t_{jk}^{23})} ^2}$ and ${\left\| {{T^{(123)}}} \right\|^2} = {\sum {(t_{ijk}^{123})} ^2}.$

Then we have:
\begin{eqnarray}\label{da1}
\frac{1}{{{d^3}}} + &&\frac{1}{{2{d^2}}}({\left\| {{T^{(1)}}} \right\|^2} + {\left\| {{T^{(2)}}} \right\|^2} + {\left\| {{T^{(3)}}} \right\|^2})\nonumber\\
 &&+ \frac{1}{{4d}}({\left\| {{T^{(12)}}} \right\|^2} + {\left\| {{T^{(13)}}} \right\|^2} + {\left\| {{T^{(23)}}} \right\|^2}) + \frac{1}{8}{\left\| {{T^{(123)}}} \right\|^2} = 1.
\end{eqnarray}

One computes that:
\begin{equation*}
{\rho _{1}} = \frac{1}{d}I + \frac{1}{2}\sum {t_i^1\lambda _i^1},{\rho _{23}} = \frac{1}{d^2}I \otimes I + \frac{1}{{2d}}(\sum {t_j^2} \lambda _j^2 \otimes I + \sum {t_k^3} I \otimes \lambda _k^3) + \frac{1}{4}\sum {t_{jk}^{23}} {\lambda _j} \otimes {\lambda _k}.
\end{equation*}
Thus we have:
\begin{equation*}
{\rm {Tr}}\rho _1^2 = \frac{1}{d} + \frac{1}{2}{\left\| {{T^{(1)}}} \right\|^2}
\end{equation*}
and
\begin{equation*}
{\rm {Tr}}\rho _{23}^2 = \frac{1}{{{d^2}}} + \frac{1}{{2d}}({\left\| {{T^{(2)}}} \right\|^2} + {\left\| {{T^{(3)}}} \right\|^2}) + \frac{1}{4}{\left\| {{T^{(23)}}} \right\|^2}.
\end{equation*}
Similarly we get:
\begin{equation*}
{\rm {Tr}}\rho _2^2 = \frac{1}{d} + \frac{1}{2}{\left\| {{T^{(2)}}} \right\|^2},
\end{equation*}
\begin{equation*}
{\rm {Tr}}\rho _{13}^2 = \frac{1}{{{d^2}}} + \frac{1}{{2d}}({\left\| {{T^{(1)}}} \right\|^2} + {\left\| {{T^{(3)}}} \right\|^2}) + \frac{1}{4}{\left\| {{T^{(13)}}} \right\|^2};
\end{equation*}
and
\begin{equation*}
{\rm {Tr}}\rho _3^2 = \frac{1}{d} + \frac{1}{2}{\left\| {{T^{(3)}}} \right\|^2},
\end{equation*}
\begin{equation*}
{\rm {Tr}}\rho _{12}^2 = \frac{1}{{{d^2}}} + \frac{1}{{2d}}({\left\| {{T^{(1)}}} \right\|^2} + {\left\| {{T^{(2)}}} \right\|^2}) + \frac{1}{4}{\left\| {{T^{(12)}}} \right\|^2}.
\end{equation*}

By noticing that we are now considering pure state $\rho  = \left| \psi  \right\rangle \left\langle \psi  \right|,$ one has
\begin{equation}
{\rm {Tr}}\rho _i^2 = {\rm {Tr}}\rho _{jk}^2
\end{equation}
for $ijk\in\{123,213,312\}$. Then we get:
\begin{eqnarray*}
&&\frac{3}{d} + \frac{1}{2}({\left\| {{T^{(1)}}} \right\|^2} + {\left\| {{T^{(2)}}} \right\|^2} + {\left\| {{T^{(3)}}} \right\|^2})\\
 &&= \frac{3}{{{d^2}}} + \frac{1}{d}({\left\| {{T^{(1)}}} \right\|^2} + {\left\| {{T^{(2)}}} \right\|^2} + {\left\| {{T^{(3)}}} \right\|^2}) + \frac{1}{4}({\left\| {{T^{(12)}}} \right\|^2} + {\left\| {{T^{(13)}}} \right\|^2} + {\left\| {{T^{(23)}}} \right\|^2}).
\end{eqnarray*}
Set $$A = {\left\| {{T^{(1)}}} \right\|^2} + {\left\| {{T^{(2)}}} \right\|^2} + {\left\| {{T^{(3)}}} \right\|^2}$$
and $$B = {\left\| {{T^{(12)}}} \right\|^2} + {\left\| {{T^{(13)}}} \right\|^2} + {\left\| {{T^{(23)}}} \right\|^2}.$$
We have:
\begin{equation}
\label{2}
\frac{1}{4}B = (\frac{1}{2} - \frac{1}{d})A + \frac{3}{d} - \frac{2}{{{d^2}}}.
\end{equation}
Substitute (\ref{2}) into (\ref{da1}), one has
\begin{equation*}
\frac{1}{{{d^3}}} + \frac{1}{{2{d^2}}}A + \frac{1}{d}\left[ {(\frac{1}{2} - \frac{1}{d})A + \frac{3}{d} - \frac{3}{{{d^2}}}} \right] + \frac{1}{8}{\left\| {{T^{(123)}}} \right\|^2} = 1.
\end{equation*}
Furthermore, we have
\begin{equation*}
\frac{1}{8}{\left\| {{T^{(123)}}} \right\|^2} = 1 - \frac{1}{{{d^3}}} - \frac{3}{{{d^2}}} + \frac{3}{{{d^3}}} - \frac{{d - 1}}{{2{d^2}}}A \le 1 + \frac{2}{{{d^3}}} - \frac{3}{{{d^2}}}.
\end{equation*}
Thus one has:
$${\left\| {{T^{(123)}}} \right\|^2} \le \frac{1}{{{d^3}}}(8{d^3} - 24d + 16).$$

Let $\rho  \in {\rm H}_1^d \otimes {\rm H}_2^d \otimes {\rm H}_3^d$ be an arbitrary mixed state with ensemble decomposition $\rho  = \sum {{p_\alpha }} \left| {{\psi _\alpha }} \right\rangle \left\langle {{\psi _\alpha }} \right|.$
We have:
\begin{equation}
{\left\| {{T^{(123)}}(\rho )} \right\|^2} = {\left\| {\sum {{p_\alpha }{T^{(123)}}(\left| {{\psi _\alpha }} \right\rangle )} } \right\|^2}\\\le {\sum {{p_\alpha }\left\| {{T^{(123)}}(\left| {{\psi _\alpha }} \right\rangle )} \right\|} ^2}\\                                                                         \le \frac{1}{{{d^3}}}(8{d^3} - 24d + 16).
\end{equation}\hfill \rule{1ex}{1ex}

\subsection{Proof of Theorem 2}

{\bf{Proof:}} We start the proof with pure state situation. Let $\rho=|\psi\ra\la\psi|$ be a pure quantum state in ${\rm H}_1^d \otimes {\rm H}_2^d \otimes {\rm H}_3^d\otimes {\rm H}_4^d$ with Bloch representation (6) in the main tex. By setting
$A=\sum_{i=1}^4||T^{(i)}||^2, B=\sum_{1\leq i<j\leq4}||T^{(ij)}||^2, C=\sum_{1\leq i<j<k\leq4}||T^{ijk}||^2$, and $D=||T^{(1234)}||^2$, one gets
\begin{eqnarray}\label{p21}
{\rm {Tr}}{(\rho^2)}=\frac{1}{d^4}+\frac{1}{2d^3}A+\frac{1}{4d^2}B+\frac{1}{8d}C+\frac{1}{16}D=1.
\end{eqnarray}

One can further computes for any $1\leq i\leq 4, 1\leq s<t\leq 4$ and $1\leq x<y<z\leq 4$ that
\begin{eqnarray*}
{\rm {Tr}}{(\rho_i^2)}&=&\frac{1}{d}+\frac{1}{2}||T^{(i)}||^2;\\
{\rm {Tr}}{(\rho_{st}^2)}&=&\frac{1}{d^2}+\frac{1}{2d}(||T^{(s)}||^2
                   +||T^{(t)}||^2)+\frac{1}{4}||T^{(st)}||^2;\\
{\rm {Tr}}{(\rho_{xyz}^2)}&=&\frac{1}{d^3}+\frac{1}{2d^2}(||T^{(x)}||^2+||T^{(y)}||^2+||T^{(z)}||^2)\\
                  &&+\frac{1}{4d}(||T^{(xy)}||^2+||T^{(xz)}||^2+||T^{(yz)}||^2)+\frac{1}{8}||T^{(xyz)}||^2.
\end{eqnarray*}

Since we are considering the pure state $\rho=|\psi\ra\la\psi|$, we have
\be\label{4e} {\rm {Tr}}{(\rho_i^2)}={\rm {Tr}}{(\rho_{jkl}^2)}\ee holds for any $ijkl\in\{1234, 2134, 3124, 4123\}$. Then by summing the equations in (\ref{4e}), we obtain
\be\label{s4e} \frac{1}{4d^2}B=\frac{2d^2-2}{d^4}+\frac{d^2-3}{4d^3}A-\frac{1}{8}C.\ee
Substituting (\ref{s4e}) into (\ref{p21}), we get
\begin{eqnarray*}
\frac{1}{16}D=1-\frac{1}{d^4}-\frac{2d^2-2}{d^4}-\frac{d^2-1}{4d^3}A-\frac{1}{16d}C\leq
1-\frac{1}{d^4}-\frac{2d^2-2}{d^4}=\frac{(d^2-1)^2}{d^4},
\end{eqnarray*}
which is just $D\leq \frac{16(d^2-1)^2}{d^4}$.

Then we consider a mixed state $\rho$ with ensemble representation
$\rho=\sum_{\alpha}p_{\alpha}|\psi_{\alpha}\ra\la \psi_{\alpha}|$, where $\sum_{\alpha}p_{\alpha}=1$. By the convexity of the Frobenius norm, one derives
$D\leq \sum_{\alpha}p_{\alpha}D_{\alpha}\leq\frac{16(d^2-1)^2}{d^4}$, which ends the proof. \hfill \rule{1ex}{1ex}

\subsection{Proof of Theorem 3}

{\bf{Proof:}} Let $|\psi\ra\in {\rm H}_1^d \otimes {\rm H}_2^d \otimes {\rm H}_3^d\otimes {\rm H}_4^d$ be pure state. Without lose of generality, one sets
$$|\psi\ra=\left\{
\begin{aligned}
|\phi_1\ra \otimes |\phi_{234}\ra,\ \ \ \ \ \ \ \ \ \ \ \ \ \ \ \ \ &&\textrm{if $|\psi\ra$ is 1-3 separable;}\\
|\psi_{12}\ra \otimes |\psi_{34}\ra,\ \ \ \ \ \ \ \ \ \ \ \ \ \ \ \ \ &&\textrm{if $|\psi\ra$ is 2-2 separable;}\\
|\xi_1\ra\otimes |\xi_2\ra \otimes |\xi_{34}\ra,\ \ \ \ \ \ \ \ \ \ \ &&\textrm{if $|\psi\ra$ is 1-1-2 separable;}\\
|\chi_1\ra \otimes |\chi_2\ra\otimes |\chi_3\ra\otimes |\chi_4\ra,\ \  &&\textrm{if $|\psi\ra$ is 1-1-1-1 separable}.
\end{aligned}\right.$$

 We have
\begin{eqnarray*}
t^{1234}_{ijkl}&=&{\rm {Tr}} (|\psi\ra\la\psi|\lambda_i\otimes\lambda_j\otimes\lambda_k\otimes\lambda_l)\\
&=&\left\{
\begin{aligned}
{\rm {Tr}} (|\phi_1\ra\la\phi_1|\lambda_i){\rm {Tr}} (|\phi_{234}\ra\la\phi_{234}|\lambda_j\otimes\lambda_k\otimes\lambda_l),\ \ \ \ \ \ \ \ \ \ \ \ \ \ \ \ \ &&\textrm{if $|\psi\ra$ is 1-3 separable;}\\
{\rm {Tr}} (|\psi_{12}\ra\la\psi_{12}|\lambda_i\otimes\lambda_j){\rm {Tr}} (|\psi_{34}\ra\la\psi_{34}|\lambda_k\otimes\lambda_l),\ \ \ \ \ \ \ \ \ \ \ \ \ \ \ \ \ &&\textrm{if $|\psi\ra$ is 2-2 separable;}\\
{\rm {Tr}} (|\xi_1\ra\la\xi_1|\lambda_i){\rm {Tr}} (|\xi_{2}\ra\la\xi_{2}|\lambda_j){\rm {Tr}}(|\xi_{34}\ra\la\xi_{34}|\lambda_k\otimes\lambda_l),\ \ \ \ \ \ \ \ \ \ \ &&\textrm{if $|\psi\ra$ is 1-1-2 separable;}\\
{\rm {Tr}} (|\chi_1\ra\la\chi_1|\lambda_i){\rm {Tr}} (|\chi_2\ra\la\chi_2|\lambda_j){\rm {Tr}} (|\chi_3\ra\la\chi_3|\lambda_k){\rm {Tr}} (|\chi_4\ra\la\chi_4|\lambda_l), &&\textrm{if $|\psi\ra$ is 1-1-1-1 separable}.
\end{aligned}\right.\\
&=&\left\{
\begin{aligned}
t^1_it^{234}_{jkl},\ \ \ \ \ \ \ \ \ \ \ \ \ \ \ \ \ &&\textrm{if $|\psi\ra$ is 1-3 separable;}\\
t^{12}_{ij}t^{34}_{kl},\ \ \ \ \ \ \ \ \ \ \ \ \ \ \ \ \ &&\textrm{if $|\psi\ra$ is 2-2 separable;}\\
t^1_it^2_jt^{34}_{kl},\ \ \ \ \ \ \ \ \ \ \ \ \ \ \ \ \ &&\textrm{if $|\psi\ra$ is 1-1-2 separable;}\\
t^1_it^2_jt^3_kt^4_l, \ \ \ \ \ \ \ \ \ \ \ \ \ \ \ \ &&\textrm{if $|\psi\ra$ is 1-1-1-1 separable}.
\end{aligned}\right.\\
\end{eqnarray*}
Thus
\begin{eqnarray*}
||T^{1234}||^2&=&\left\{
\begin{aligned}
||T^{1}||^2||T^{234}||^2,\ \ \ \ \ \ \ \ \ \ \ \ \ \ \ \ \ &&\textrm{if $|\psi\ra$ is 1-3 separable;}\\
||T^{12}||^2||T^{34}||^2,\ \ \ \ \ \ \ \ \ \ \ \ \ \ \ \ \ &&\textrm{if $|\psi\ra$ is 2-2 separable;}\\
||T^{1}||^2||T^{2}||^2||T^{34}||^2,\ \ \ \ \ \ \ \ \ \ \ &&\textrm{if $|\psi\ra$ is 1-1-2 separable;}\\
||T^{1}||^2||T^{2}||^2||T^{3}||^2||T^{4}||^2, \ \ \ \ &&\textrm{if $|\psi\ra$ is 1-1-1-1 separable}.
\end{aligned}\right.\\
&\le&\left\{
\begin{aligned}
\frac{16}{d^4}((d-1)(d^3-3d+2)),&&\textrm{if $|\psi\ra$ is 1-3 separable;}\\
\frac{16}{d^4}(d^2-1)^2,\ \ \ \ \ \ \ \ \ \ \ \ \ \ \ \ \ &&\textrm{if $|\psi\ra$ is 2-2 separable;}\\
\frac{16}{d^4}(d^2-1)(d-1)^2,\ \ \ \ \ \ \ \ &&\textrm{if $|\psi\ra$ is 1-1-2 separable;}\\
\frac{16}{d^4}(d-1)^4,\ \ \ \ \ \ \ \ \ \ \ \ \ \ \ \ \ \ \ &&\textrm{if $|\psi\ra$ is 1-1-1-1 separable}.
\end{aligned}
\right.
\end{eqnarray*}

Then for any mixed state $\rho\in {\rm H}_1^d \otimes {\rm H}_2^d \otimes {\rm H}_3^d\otimes {\rm H}_4^d$, one has
\begin{eqnarray*}
\| T^{(1234)}\|^2&=&\|\sum_kp_kT_k^{(1234)}\|^2\leq\sum_kp_k\| T_k^{(1234)}\|^2\\
&&\leq\left\{
\begin{aligned}
\frac{16}{d^4}((d-1)(d^3-3d+2)),& & \textrm{if $\rho$ is 1-3 separable;}\\
\frac{16}{d^4}(d^2-1)^2,\ \ \ \ \ \ \ \ \ \ \ \ \ \ \ \ \ & & \textrm{if $\rho$ is 2-2 separable;}\\
\frac{16}{d^4}(d^2-1)(d-1)^2,\ \ \ \ \ \ \ \ & & \textrm{if $\rho$ is 1-1-2 separable;}\\
\frac{16}{d^4}(d-1)^4,\ \ \ \ \ \ \ \ \ \ \ \ \ \ \ \ \ \ \ & & \textrm{if $\rho$ is 1-1-1-1 separable},
\end{aligned}
\right.\end{eqnarray*}
which ends the proof.\hfill \rule{1ex}{1ex}

\end{document}